\documentstyle[psfig]{l-aa}

\newcommand{\ieff}{\mbox{{\scriptsize eff}}}
\newcommand{\ifit}{\mbox{{\scriptsize fit}}}
\newcommand{\iobserved}{\mbox{{\scriptsize observed}}}
\newcommand{\ie}{\mbox{{\scriptsize e}}}
\newcommand{\idavis}{\mbox{{\scriptsize Davis}}}
\newcommand{\itw}{\mbox{{\scriptsize this work}}}
\newcommand{\ibrett}{\mbox{{\scriptsize Brett}}}

\begin{document}

   \thesaurus{07         
	     (02.13.4;   
              08.01.3;   
	      08.16.4;   
	      08.22.3)}  
   \title{Near-infrared narrow-band photometry of M--giant and Mira 
          stars: models meet observations}


   \author{R.\ Alvarez
	   \inst{1}
	   and B.\ Plez
	   \inst{2,}\inst{3}
	  }

   \offprints{alvarez@graal.univ-montp2.fr}

   \institute{GRAAL,
	      Universit\'e Montpellier II,
              UPRESA 5024/CNRS,
	      F--34095 Montpellier Cedex 05, France
	      \and
	      Astronomiska Observatoriet, 
              box 515,
              S--751 20 Uppsala, Sweden
	      \and
              Atomspektroskopi,
	      Fysiska Institution, 
              box 118,
              S--221 00 Lund, Sweden,
              plez@fysik.lu.se
              }

   \date{Received 20 June 1997; Accepted 19 September 1997}

   \maketitle

   \markboth{Alvarez \& Plez: NIR photometry of M--giants and Miras}
            {Models meet observations}

   \begin{abstract}
    From near-infrared, narrow-band photometry of 256 oxygen-rich Mira
    variables we obtain evidence about the loops that these stars
    follow in colour--colour diagrams. We also find a phase lag between
    indices related to molecular band-strength of titanium oxide and 
    vanadium oxide.
    We compute colours for normal M--giants and Miras using 
    hydrostatic and hydrodynamic model atmospheres and very extensive 
    up-to-date line lists.
    Normal M--giants colours are well reproduced, reaching a high 
    quantitative agreement with observations for spectral types earlier
    than M7.
    The out-of-phase variations of the various spectral features of
    Miras are also acceptably reproduced, despite limitations
    in the modelling. This enables us to confirm that the phase lag
    phenomenon results from the propagation of perturbations in the 
    extended atmosphere. It opens new perspectives in the spectral 
    modelling of Miras.

   \end{abstract}


\section{Introduction}

Red giants and long-period variables provide crucial information on 
stellar and galactic evolution. It is thus essential to 
be able to interpret their spectra and photometry in terms of stellar 
parameters (effective temperature, chemical composition, mass, etc) 
and physical processes (presence of shocks, of a wind, etc).
In spite of the difficulty to model these stars due to the
many complications arising from low temperatures (molecule and dust
formation) and the often very extended atmosphere, important 
progress in the modelling of static (Plez et al.\ 1992; see also the 
review by Plez 1997) and dynamic (Bowen 1988; Fleischer et al.\ 1992; 
H\"ofner \& Dorfi 1997) cool star atmospheres have been achieved in 
the last years.\\
Confrontation with observations is the natural aim of any modelling
attempt. Checking the validity of the various models' assumptions and
inputs is obviously essential.
Some recent investigations have dealt with high resolution 
spectroscopic observations (e.g.\ Plez et al.\ 1993) while others
concentrated on photometry (Bessell et al.\ 1996). 
They demonstrate that model atmospheres and synthetic spectra are now
reaching a high quantitative agreement with observations, due to the
recent great improvement in opacity completeness and accuracy.\\
The near-infrared narrow-band photometry of Lockwood (1972), described
in Sect.~2, has various qualities which make it very suitable
for the study of M--giants and Mira variables.
It consists of narrow-band filters around 1 $\mu$m, a spectral region 
where very cool stars emit most of their energy and where, 
unfortunately, very few observed spectra are available.
The narrow-band filters used allow a precise study of absorption
features of molecules dominating the optical spectrum of cool stars,
titanium oxide (TiO) and vanadium oxide (VO), for which new data have 
been recently published. 
Furthermore,  Lockwood's data represent one of the very few set of 
regular observations on whole cycles of a large number of 
Miras. The study of the effects of pulsation on absorption features is 
thus possible. Observational evidence for a phase lag between indices
related to TiO and VO absortion features is given (Sect.~3).
In Sect.~4, synthetic colours from static models are computed 
and compared to M--giant colours. Thanks to the updated millions 
of spectral lines taken into account, a remarkable agreement is 
obtained. In Sect.~5 we compute LTE radiative transfer in hydrodynamic 
models that satisfactorily reproduce the colour variations and the 
phase lag for Mira variables. 

\section{The data}

\subsection{Lockwood's photometric system}

Lockwood (1972; hereafter L72) observed 292 M-- and S--type 
Mira variables from 1969 to 1971 at the Kitt Peak National 
Observatory, with a five-colour narrow-band photometric system based
on Wing's 27 colours system (Wing 1967). 
He obtained 1795 individual sets of five-colour measurements, at 
different phases, and, for some stars, during several cycles. 
The narrow-bands were chosen to measure depths of molecular band 
heads of TiO and VO, or regions relatively free of molecular 
absorption. 
Table~1 lists the properties of the five-colour system. 
The 78 and 88 filters were used to measure several bands of TiO. 
The VO absorption was measured by the 105 filter. The 87 and 
104 filters were expected to match 'continuum' regions.
However, it appeared to Lockwood that the 87 filter becomes 
contaminated by TiO bands in the later M stars.
In L72 are tabulated the measurements of the 104 magnitudes and 
of the four colours: 78$-$87, 88$-$87, 87$-$104, and 105$-$104.
The median standard errors were 0.012 mag for 104 and 0.006 mag for 
the colours.

\begin{table}
\caption[]{Properties of Lockwood's (1972) five-colour system}
\begin{flushleft}
\begin{tabular}{llll}
\hline\noalign{\smallskip}
Filter      & Peak       & Half-power & Feature \\
designation & wavelength & bandwidth  & (Lockwood 1972) \\
\noalign{\smallskip}
\hline\noalign{\smallskip}
 78 &  7818 \AA &  90 \AA & TiO \\
 87 &  8777 \AA &  82 \AA & Continuum (+TiO) \\
 88 &  8884 \AA & 114 \AA & TiO \\
104 & 10351 \AA & 125 \AA & Continuum \\
105 & 10506 \AA & 100 \AA & VO \\
\noalign{\smallskip}
\hline
\end{tabular}
\end{flushleft}
\end{table}

Due to the regularity of the observations, this set of data allows a 
precise study of the behaviour of TiO and VO absorption with phase.
For homogeneity reasons, only the oxygen-rich Miras have been 
subsequently considered. 
They represent a sample of 256 stars for which 1501 sets of 
five-colour measurements were obtained at several phases. The period 
distribution of the sample corresponds perfectly to the one obtained 
for the M--Miras listed in the General Catalogue of Variable Stars 
(GCVS; Kholopov 1985, 1987). The stars were observed preferably near 
maximum and minimum of light curves.

\subsection{Molecular band-strength indices}

In order to study the behaviour of TiO and VO absorption features with 
phase, we have considered four indices:
\begin{description}
   \item 78$-$88: this index is proportional to the ratio between two 
   different band heads of TiO. It is a good effective temperature 
   indicator for temperatures above 3000~K and thus was prefered 
   to 88$-$87 originally given in L72. For lower temperatures, the 
   TiO bands begin to saturate.
   \item 78$-$87: the 87 and 88 filters overlap slightly. 
   As a consequence, the 78$-$88 and 78$-$87 indices are very similar. 
   The absolute differences are always less than 0.2 mag.
   \item 87$-$104: Lockwood originally hoped to obtain a colour 
   temperature determination from this index as both filters were 
   thought to measure regions free of strong molecular absorption. But 
   filter 87 proves to be contaminated at least by TiO bands.
   \item 105$-$104: it is related to VO bands. This index depends 
   strongly on effective temperature below 3000~K where the VO 
   molecular bands begin to appear.
\end{description}
The first three indices can be considered to measure mainly TiO 
features (although we will show below that VO bands are also present 
in these passbands) and 105$-$104 may be seen as a pure VO index. The 
indices 78$-$88 and 105$-$104 complement each other very well as 
temperature indicators. They have been used to determine effective 
temperature for a large number of Miras (Alvarez \& Mennessier 1997).
The four colours are each defined so as to increase linearly with
decreasing effective temperature during a cycle.

\section{Variations of TiO and VO absorption features during a cycle.
         Observational evidence for phase lag}

\subsection{Phase--colour diagrams}

The phases of all sets of measurements are given in L72. They 
were determined from the mean periods given in the GCVS and the AAVSO 
annual predictions. 
Phase zero marks the maximum of the visual light curve.
Figure 1 shows the four indices calculated from the entire data 
(1501 measurements) plotted against the phase.
The curves represent the running average computed for 100 points on
a 1/16th cycle interval.
The data in the phase interval [\,0.5\,;\,1.0\,] are reproduced in the 
interval [\,-0.5\,;\,0.0\,].
Arrows indicate the extremes of each curve.

\begin{figure*}
  \centerline{\psfig{figure=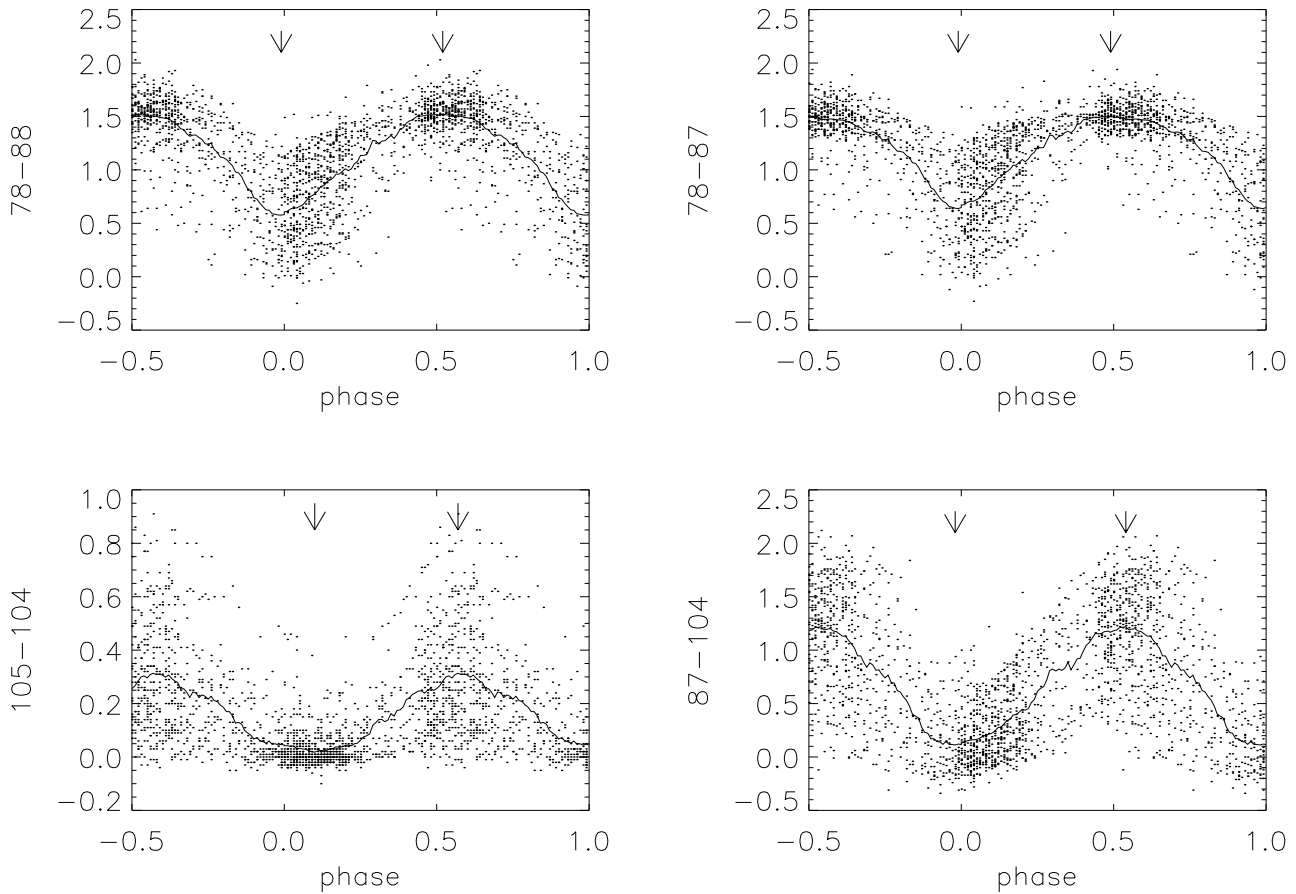,width=18cm}}
  \caption[]{The four colour indices as a function of phase 
   (1501 measurements for 256 O-rich Miras). 
   The curves are the running average values and the arrows indicate 
   the extreme values}
\end{figure*}

Such a plot has of course no real physical significance since the data 
are composed of observations of 256 different Miras, which exhibit a 
large spread of periods and mean spectral types.
Nevertheless, it gives interesting insights into the global behaviour 
of the indices with phase.
As expected, the colours reach a minimum around phase zero when 
the effective temperature is highest and reach a maximum around 
phase 0.5, at minimum light.\\
The important feature is the phase lag that exists between the 
105$-$104 index and the other indices. The VO index has its minimum 
value at phase $\varphi=0.1$ or so, whereas the indices 
78$-$88, 78$-$87, 87$-$104 
have their minimum slightly before phase zero.
The same phase shift of about 0.1 is also found between the maxima: 
it occurs at $\varphi=0.6$ for 105$-$104, and around $\varphi=0.5$ for 
the other colours. 
Note that the dispersion around the mean value is minimum near phase 
zero for 87$-$104 and 105$-$104 and maximum near phase 0.5, whereas 
the opposite is true of the two other indices. This is explained by 
the sensitivity, or the lack thereof, of the various indices in the 
effective temperature range above or below 3000~K (see Fig.~7). 
As a consequence, the positions of the maxima of the TiO indices 
(78$-$87, 78$-$88) which saturate around 1.8, are not easy to locate 
accurately. 
The VO index (105$-$104) has values clearly greater than zero only 
near the minimum light. Thus, the presence or not of a phase lag is 
not totally obvious in Fig.~1. 
Colour--colour diagrams enable us to confirm that feature.

\subsection{Colour--colour diagrams}

Figures 2 and 3 display the average colour--colour variation with 
phase for 78$-$88/105$-$104, and 78$-$87/87$-$104 respectively.
The average curves of Fig.~1 are simply reported in those
colour--colour diagrams. The filled circles mark phase zero and
the arrows indicate the direction of variation during the cycle.\\
In the 78$-$88/105$-$104 diagram (Fig.~2), a loop is clearly described 
during the cycle. To a single 78$-$88 value two different 105$-$104 
values corresponds, depending on whether the Mira is rising or 
declining. This loop confirms that a phase lag exists between 
78$-$88 and 105$-$104.
On the other hand, there is no loop in the 78$-$87/87$-$104 diagram 
(Fig.~3). It demonstrates the absence of phase shift and thus the 
occurence of extremes at more or less the same phase, as can be seen 
in Fig.~1. 
Similarly there is no phase lag between 78$-$88 and 78$-$87. Only 
105$-$104 is out-of-phase by an amount of 0.1.

\begin{figure}
  \centerline{\psfig{figure=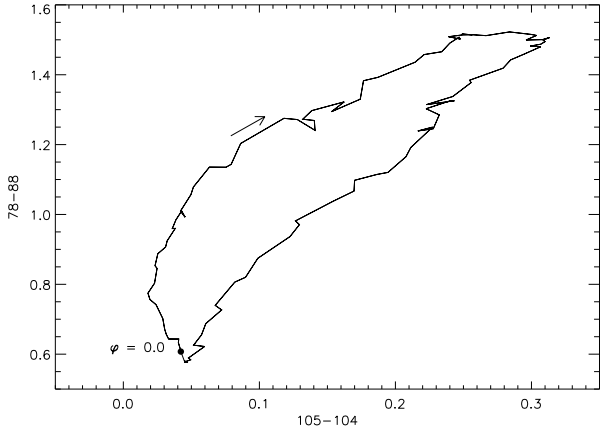,width=3.4in}}
  \caption[]{78$-$88/105$-$104 diagram. The curve represent the running 
   average values shown in Fig.~1. The filled circle marks the phase zero 
   and the arrow indicates the direction of variation during the cycle}
\end{figure}

\begin{figure}[t]
  \centerline{\psfig{figure=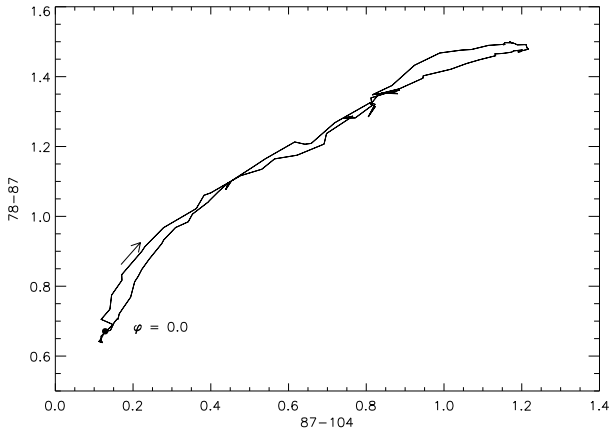,width=3.4in}}
  \caption[]{Same as Fig.~2 for 78$-$87 and 87$-$104.}
\end{figure}

The presence (absence) of loop in the 78$-$88/105$-$104 
(78$-$87/87$-$104) diagram is also verified when only a part of the 
sample of Miras is considered. Early and late spectral type Miras or 
short and long period sub-samples have been tried. The apparent phase 
shift and the presence of a loop in a colour--colour diagram which is 
its counterpart are independent of period and spectral type.
The loop for shorter periods (higher $T_{\ieff}$) is at smaller colour
indices than that for longer periods (lower $T_{\ieff}$).

\subsection{Individual phase shifts}

\subsubsection{Method}

The results described above concern only mean values. The trends 
revealed by this {\it global approach} should be verified by 
considering the individual sets of observations. 
In the following, we adopted the complementary {\it individual approach} 
to determine the possible phase shifts.\\ 
For each Mira with a number of observations at least equal to six 
(93 objects), we fitted the four indices as a function of phase 
$\varphi$ by sine curves:
\begin{equation}
   c_{i} = a_{0} + a_{1} \sin [2 \pi (\varphi - \varphi_{0})]
\end{equation}
where $c_{i}$ is one of the four colours; $a_{0}$, $a_{1}$ and 
$\varphi_{0}$ are the three free parameters.\\
We only used the Miras which have observed colours distributed in a 
phase range larger than 0.3 in order to properly fit the observations. 
We excluded the stars for which:
\begin{equation}
   \max (c_{i})_{\ifit}  >  \max (c_{i})_{\iobserved} + 0.4
\end{equation}
\begin{equation}
   \min (c_{i})_{\ifit}  <  \min (c_{i})_{\iobserved} - 0.4
\end{equation}
For these stars, the extremes given by the fit are too distant from
the observations and thus are less reliable.
Finally, we rejected the Miras with:
\begin{equation}
   \max (105-104)_{\ifit}  <  0.05 \\
\end{equation}
as it means that these stars show no appearance of VO absorption bands 
during the whole cycle (the temperature is always too high) and the 
fits have no meaning in these cases. There is no such problem with the 
other colours.\\
Among the 93 Miras with a number of observations greater than 5, 76 
finally remained. 
The phase shift between two indices is simply:
\begin{equation}
   \delta\varphi = \varphi_{0}(c_{i}) - \varphi_{0}(c_{j}) \\
\end{equation}

\subsubsection{Results}

Figure 4 presents the histogram of phase shifts between the 
(78$-$88, 105$-$104) indices. A majority of Miras clearly displays a 
phase lag. The distribution is almost symmetrical, the mean value 
being 0.08, with a sigma of 0.07. 
The variation of the 105$-$104 index lags behind that of 78$-$88, as 
it is observed in Fig.~1. 
The individual phase lag values are not correlated with period or 
spectral type.
The histogram for the (78$-$87, 87$-$104) indices is shown in Fig.~5.
No global phase lag is apparent as in Fig.~4. 
The calculated mean value is 0.01 with a sigma of 0.07.\\
The fit of the observations by sine curves is of course not really 
accurate and the individual phase lags determined with this method are 
not reliable, but both distributions provide an excellent statistical
confirmation of what was concluded from the previous global approach:
the 105$-$104 index appears out-of-phase with the other indices by an 
amount of almost 0.1, which corresponds to more or less 30 days for
a typical Mira.

\begin{figure}
  \centerline{\psfig{figure=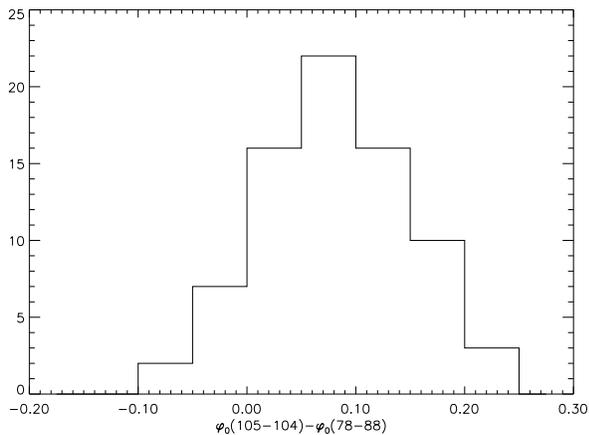,width=3.4in}}
  \caption[]{Histogram of phase shifts: 
   $\varphi_{0}(105-104)$ minus $\varphi_{0}(78-88)$}
\end{figure}

\begin{figure}
  \centerline{\psfig{figure=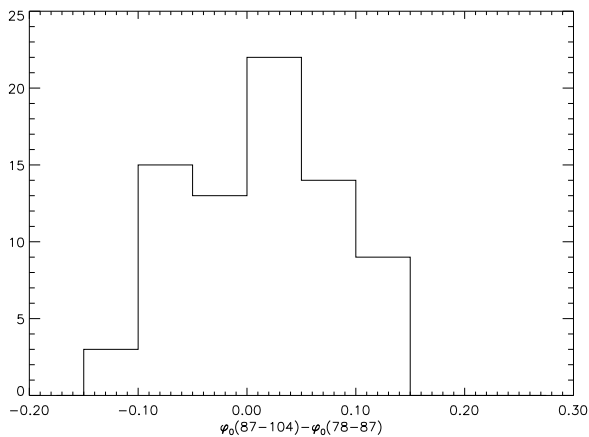,width=3.4in}}
  \caption[]{Histogram of phase shifts: 
   $\varphi_{0}(87-104)$ minus $\varphi_{0}(78-87)$}
\end{figure}

\subsection{Comments on the phase lag}

Lockwood (1972) already noted that Miras perform large loops in a
colour/spectral-type diagram. He pointed out that the mean phase of 
maximum light at 1.04 $\mu$m occurs at 0.11, whereas at V it occurs at
phase zero by definition. Mean minimum light at 1.04 $\mu$m is reached 
almost simultaneously with mean minimum visual light, although the 
individual phases of these events often differ by 0.1 cycle or 
more.\\
We have shown that the phase lag and the loop phenomenon are also 
displayed by the particular near infrared indices considered here. 
More precisely, the phase shift is between the VO index on one hand 
and indices mainly related to TiO on the other hand.
The phase lag between changes in the various bands which are dominated 
by different molecular absorption (TiO and VO) may be explained by the 
fact that these bands are formed at various depths in the atmosphere. 
What we see is a consequence of the shocks (created by the pulsation) 
running through the atmosphere.\\
Important progress in nonlinear stellar pulsation models has been 
made in the last years (Feuchtinger \& Dorfi 1994, 1996a, 1996b; 
Bessell et al.\ 1996; Ya'ari \& Tuchman 1996). 
Dynamical atmosphere models have also been successfully developed 
recently (e.g.\ Bowen 1988; Fleischer et al.\ 1992; 
H\"ofner \& Dorfi 1997): the effects of pulsation generated in the 
interiors are simulated by periodic mechanical driving at the inner 
boundary (piston model). 
All these numerical simulations show that the propagation of 
successive shock fronts in the atmosphere creates a complex 
temperature, density, and velocity stratification. 
Parts of the atmosphere are heated up while others cool down. The run 
of temperature is often far from the smooth decrease found in static 
LTE models. 
As a consequence, during a fraction of the cycle, the VO molecular 
lines can still be building up strength, as e.g.\ VO molecules are 
forming (the VO index increases), while the TiO lines have started 
to weaken (TiO indices decrease): a loop then appears in the 
corresponding colour--colour diagram.
We will try to investigate this hypothesis in Sect.~5 by using a 
set of dynamical atmosphere models to compute the colours variation
during a cycle, but this requires that we are first able to 
well reproduce Lockwood's colours for non-variable M--giants.
This is the aim of the following section. 

\section{M--giant standards and hydrostatic model colours}

The first step in any attempt to investigate the origin of the phase 
lag is to correctly reproduce Lockwood's colours for M--giants. 
It is now possible thanks to important progress in the modelling of 
red giant atmospheres achieved in the last 
years (Plez et al.\ 1992, hereafter PBN92; see also the review by 
Plez 1997).

\subsection{The synthetic spectrum program}

We computed synthetic spectra with TurboSpectrum, an enhanced version 
of the Spectrum package developed at Uppsala observatory and
used in previous investigations of AGB stars (e.g.\ Plez et al.\ 1993). 
Input model atmospheres were calculated with SOSMARCS (PBN92), one of 
the latest developments in the MARCS suite of programs initiated by
Gustafsson et al.\ (1975). SOSMARCS generates spherically symmetric, 
hydrostatic, flux constant (convective + radiative) models in LTE. 
It is especially designed for cool star atmospheric conditions. It 
features a full Opacity Sampling treatment of all opacities with 
10912 sampling points between 990 and 125000~\AA. 
TurboSpectrum is based for a large part on the same ensemble of 
routines (e.g.\ evaluation of continuous opacities and solution of the 
transfer equations). Any kind of temperature--pressure stratification 
may be used and the execution time is relatively short even when 
millions of lines are taken into account.
It offers the possibility for instance to test complex profiles which
can mimic pulsating atmospheres, and perform extensive calculations 
with various abundances and/or line lists.

\subsection{The atomic and molecular data}

We included for the spectrum computations atomic line data 
from N to Ni (Kurucz 1992). 
We also included line lists for 
the TiO $\alpha$, $\beta$, $\gamma$, $\gamma$', $\delta$, 
$\epsilon$, $\phi$ and a--f electronic transition systems, 
the VO A--X and B--X electronic transition systems,
and a newly calculated line list for H$_{2}$O (J\o rgensen, private
communication).
The absorption systems of $^{12}$C$^{14}$N and $^{13}$C$^{14}$N 
(J\o rgensen \& Larsson 1990) are also taken into account, 
with a $^{12}$C/$^{13}$C ratio equal to 19, which is a typical 
value after the first dredge-up (Smith 1990).

Most of the TiO and VO band strengths were updated since PBN92.
In PBN92 the electronic transition momenta ($R_{\ie}^2$) 
of Davis et al.\ (1986) were adopted for the TiO $\alpha$, $\beta$, 
$\gamma$, $\gamma$', $\delta$ and $\phi$ systems, supplemented by 
Brett's astrophysical calibration for the TiO $\epsilon$ (Brett 1990).
In subsequent works (e.g.\ Plez et al.\ 1993) the lifetime 
measurements of Doverst\aa l \& Weijnitz (1992) replaced some of 
the earlier values.
Recently Hedgecock and coworkers (1995) performed lifetime 
measurements with a better accuracy and for a larger number of 
electronic states than previously for the $\alpha$, $\beta$, 
$\gamma$ and $\gamma$' systems. These new lifetime values are 
significantly different from the older ones. We decided to adopt 
them in our investigation. 
We generated also a line list for the a--f system using 
the lifetime given by Hedgecock et al.\ 
(1995, experimental), and Schamps et al.\ (1992, theoretical) which 
agree within 5\%. This system absorbs mostly in the V band region and 
has no impact on the present calculations.
For the TiO $\phi$ (Hedgecock et al.\ did not measure the
corresponding lifetime) and the $\epsilon$ (the same authors provide 
only an upper limit of the strength) systems, the results of the 
recent ab initio calculations performed by Langhoff (1997) are 
adopted. His calculations for the other bands are in good agreement 
with the Hedgecock et al.\ values.   
We included all 5 stable isotopes of Ti with terrestrial abundance 
ratios in the computation of the line list.\\
The VO A--X and B--X band strengths in PBN92 result from the 
astrophysical calibration by Brett (1990) and are not expected 
to be very reliable especially as they were calibrated using outdated 
mean opacity models. 
Recently, Karlsson et al.\ (1997) measured the lifetimes of the A, B 
and C electronic states. We included their values in our line list 
which was also revised using Cheung et al.\ (1982a, 1982b, 1994),
and Adam et al.\ (1995). The new line positions are shifted by up to 
65~cm$^{-1}$ ($-$40~\AA) in the B--X system, and 25~cm$^{-1}$
in the A--X system compared to PBN92. The line-strengths are scaled 
by a factor 0.51, 0.28, and 0.72 for the A--X, B--X, respective C--X 
transitions relative to PBN92's values.

We summarize the $R_{\ie}^2$ and the electronic oscillator strength 
($f_{\ie}$-values) we used in Table~2 and~3 for the TiO and VO
systems respectively. 
Like Larsson (1983), we prefer to avoid the discussion of 
$f_{\ie}$-values which lack a sound and unique definition in molecular 
spectra. We provide them anyway as a guide for comparison with other 
works (e.g.\ J\o rgensen 1994). 
We define $f_{\ie}$ through:
$f_{\ie} = 3.038 \times 10^{-6} \times \overline{\sigma}_{00}
\times R_{\ie}^2 \times 
\frac{2-\delta_{0,\Lambda\prime+\Lambda\prime\prime}}
{2-\delta_{0,\Lambda\prime\prime}}$,
where $\overline{\sigma}_{00}$ is some average wavenumber for the 
(0--0) band of the transition system, $\prime$ and $\prime\prime$ 
denote the upper and lower state respectively, and $\Lambda$ is the 
projection of the electron orbital angular momentum (0 for $\Sigma$ 
states, 1 for $\Pi$ states etc); see Larsson for more details.

\begin{table}
\begin{minipage}{8.5cm}
\caption[]{Adopted electronic transition momenta for the TiO systems}
\begin{flushleft}
\begin{tabular}{lllll}
\hline\noalign{\smallskip}
System  &  $R_{\ie,\idavis}^2$  &  $R_{\rm e,\itw}^2$  & 
resulting       &  $\frac{R_{\ie,\itw}^2}{R_{\ie,\idavis}^2}$  \\
        &                       &                      &          
$f_{\ie}$-value &                                               \\
\noalign{\smallskip}
\hline\noalign{\smallskip}
$\alpha$   & 3.0    & 1.8\footnote[1]{based on lifetime 
                          measurements (Hedgecock et al.\ 1995)}
           & 0.106  & 0.6         \\
$\beta$    & 5.14   & 2.3$^{a}$   
           & 0.125  & 0.45        \\
$\gamma$   & 3.6    & 1.84$^{a}$    
           & 0.0786 & 0.51        \\
$\gamma$'  & 2.8    & 1.9$^{a}$    
           & 0.0935 & 0.68        \\
$\delta$   & 1.4    & 1.4           
           & 0.048  & 1.0         \\
$\epsilon$ & --     & 0.064\footnote[2]{based on lifetime 
                            calculations (Langhoff 1997)} 
           & 0.0023 & --          \\
$\phi$     & 0.9    & 0.32 $^{b}$        
           & 0.0178 & 0.36        \\
a--f       & --     & 1.69$^{a}$    
           & 0.098  & --          \\
\noalign{\smallskip}
\hline
\end{tabular}
\end{flushleft}
\end{minipage}
\end{table}

\begin{table}
\begin{minipage}{8.5cm}
\caption[]{Adopted electronic transition momenta for the VO systems}
\begin{flushleft}
\begin{tabular}{lllll}
\hline\noalign{\smallskip}
System & $R_{\ie,\ibrett}^2$\footnote[1]{astrophysical 
                             calibration by Brett (1990)}
       & $R_{\ie,\itw}^2$\footnote[2]{based on lifetime 
                          measurements (Karlsson et al.\ 1997)} \\
\noalign{\smallskip}
\hline\noalign{\smallskip}
A--X   &  0.207  & 0.106  \\
B--X   &  2.590  & 0.735  \\
\noalign{\smallskip}
\hline
\end{tabular}
\end{flushleft}
\end{minipage}
\end{table}

As shown in Table~2 and~3, almost all TiO and VO band strengths have 
been obtained from recent laboratory measurements or calculations. 
Apart from the $\delta$ system, all bands have been updated since 
PBN92.
The choice to keep the Davis et al.\ value for the $\delta$ system
comes from comparisons between observed and calculated spectra for a 
sample of M giants and dwarfs. Adopting Langhoff's lifetime for 
$b^1\Pi$ and branching ratios for the $\delta$ and $\phi$ transitions 
leads to obviously too strong $\delta$ bands. We could not find an 
explanation for this mismatch and decided therefore to keep the 
seemingly better value from Davis et al.\ (1986), which also matches 
Ramsey's (1981) measurements of the $\delta$(0--0) bandhead at 
8160~\AA\ rather well down to $T_{\ieff} \sim$ 3400~K. Adopting 
Langhoff's value, which is twice as large, results in a calculated
Ramsey's $D_{8860}$ index systematically too large by 0.1 to 0.14.
This is however puzzling and will be further investigated 
by Plez et al.\ (in preparation) in their discussion of the new large 
grid of Uppsala models.\\
Davis et al.\ (1986) used the lifetime of the upper state of a 
transition in the $\beta$ system measured by Feinberg \& Davis (1977)
to put their measurement of relative transition momenta on an absolute 
scale. Our revision of the $\beta$ system strength, following 
Hedgecock et al.\ lifetime measurement, amounts to a downward scaling 
by a factor 0.45 of Davis et al.\ value. For the other bands quoted by 
Davis et al.\ the factor is between 0.36 and 0.68 (see Table~2).

\subsection{Influence of the various molecular bands on colours}

We first computed individual spectra for each absorption band
system in order to establish its particular influence on Lockwood's 
passbands.
Table~4 presents the systems which are really dominant in each band 
and those for which the influence is less pronounced but still 
measurable. The systems that are not listed have no significant 
influence.\\
The most prominent features measured by the filters are the TiO 
$\gamma$ and $\delta$ systems, and the VO A--X and B--X systems. 
A noticeable result is the relatively important influence of VO 
absorption on the 87 and 88 filters. 
Furthermore, when the $T_{\ieff}$ reaches values lower than 2800~K, 
the features due to the VO B--X system become even stronger than 
those due to TiO bands in our models. This effect was not expected 
by Lockwood. 
Concerning the continuum points, the 104 filter can effectively be 
considered as a good measurement of a region relatively free of 
strong molecular absorption; this is certainly not the case for 
the 87 filter.
The CN and H$_2$O bands and the atomic lines have no significant 
effect on the emergent fluxes in the filter passbands.\\
The Lockwood colours, as they measure the influence of a limited 
number of molecular band systems as shown in Table~4, constitute an 
ideal opportunity to check the calibration of some of those TiO and 
VO bands and/or the accuracy of the models and the synthetic spectra.

\begin{table}
\caption[]{Dominant molecular features in each band of Lockwood's 
           photometric system}
\begin{flushleft}
\begin{tabular}{lll}
\hline\noalign{\smallskip}
Filter  &  Dominant systems  &  Less influent systems  \\
\noalign{\smallskip}
\hline\noalign{\smallskip}
 78 & TiO~$\gamma$    
    & VO~B--X, TiO~$\gamma$', $\delta$ and $\epsilon$ \\
 87 & TiO~$\gamma$ and VO~B--X                         
    & TiO~$\epsilon$ \\
 88 & TiO~$\delta$, $\gamma$ and VO~B--X   
    & TiO~$\epsilon$ and VO~A--X \\
104 & --                
    & TiO~$\phi$, $\delta$, $\epsilon$ and $\gamma$ \\
105 & VO~A--X                        
    & TiO~$\epsilon$, $\gamma$, $\phi$ \\
\noalign{\smallskip}
\hline
\end{tabular}
\end{flushleft}
\end{table}

\subsection{Comparison to observations}

Lockwood has observed 61 standard red giants with spectral types 
from K5 to M8 in the same five-colour system as the Mira variables. 
In order to compare these observations with synthetic colours, we 
used fifteen SOSMARCS models representing a sequence of red giants 
with solar composition, a stellar mass of 1.5 $\cal M_{\odot}$ and 
$T_{\ieff}$ ranging from 2500 to 3900~K. The surface-gravities
$\log g$ vary from 1.60 at $T_{\ieff}=3900$~K to $-0.50$ at
$T_{\ieff}=2500$~K. 
These models were used by Fluks et al. (1994) and are further 
described there. The emergent fluxes were computed between 7500 and 
11000~\AA\ with a resolution of 0.2~\AA. Changing the sampling to 
0.8~\AA\ changes the colours by at most a few hundredth of a 
magnitude.
Figure~6 shows an example of a synthetic spectrum ($T_{\ieff}=3500$~K).
The dotted line is part of an optical spectrum of HD 123657 (BY Boo), 
a M5--giant, obtained by Serote Roos et al.\ (1996) at a resolution 
of 1.25~\AA\ using the Aurelie spectrograph, attached to the OHP 
1.52m telescope. The region covered ends at 8920~\AA. 
The agreement is very good.

\begin{figure}
  \centerline{\psfig{figure=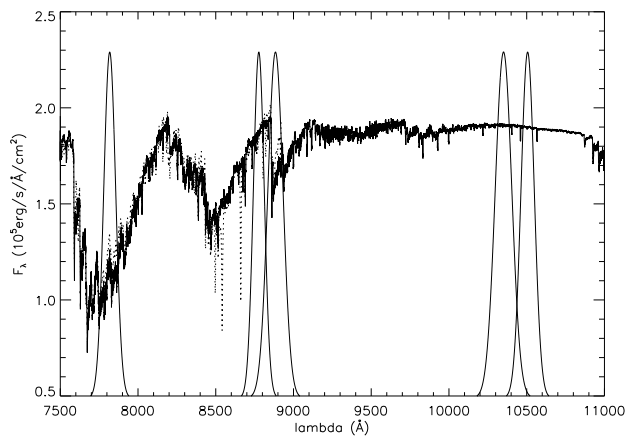,width=3.4in}}
  \caption[]{Synthetic spectrum with $T_{\ieff}=3500$~K, 
   $\log g$=0.9, $\cal M$=1.5 $\cal M_{\odot}$.
   The dotted line is part of an optical spectrum of HD 123657, 
   a M5--giant, from Serote Roos et al.\ (1996). The CaII triplet 
   lines are not included in the synthetic spectrum. The filter 
   transmission curves used in this work are superimposed (full line)}
\end{figure}

Lockwood's filters of Table~1 were applied to the fifteen 
spectra. The zero point calibration is based on a spectrum of Vega 
(Dreiling \& Bell 1980) as used by Bessell \& Brett (1988). 
An effective temperature has been assigned to each standard giant 
using the spectral types given in L72, and the spectral 
type--$T_{\ieff}$ relation determined by Fluks et al.\ (1994).
The overall good agreement of the synthetic and observed colours is 
demonstrated in Fig.~7. 

\begin{figure*}
  \centerline{\psfig{figure=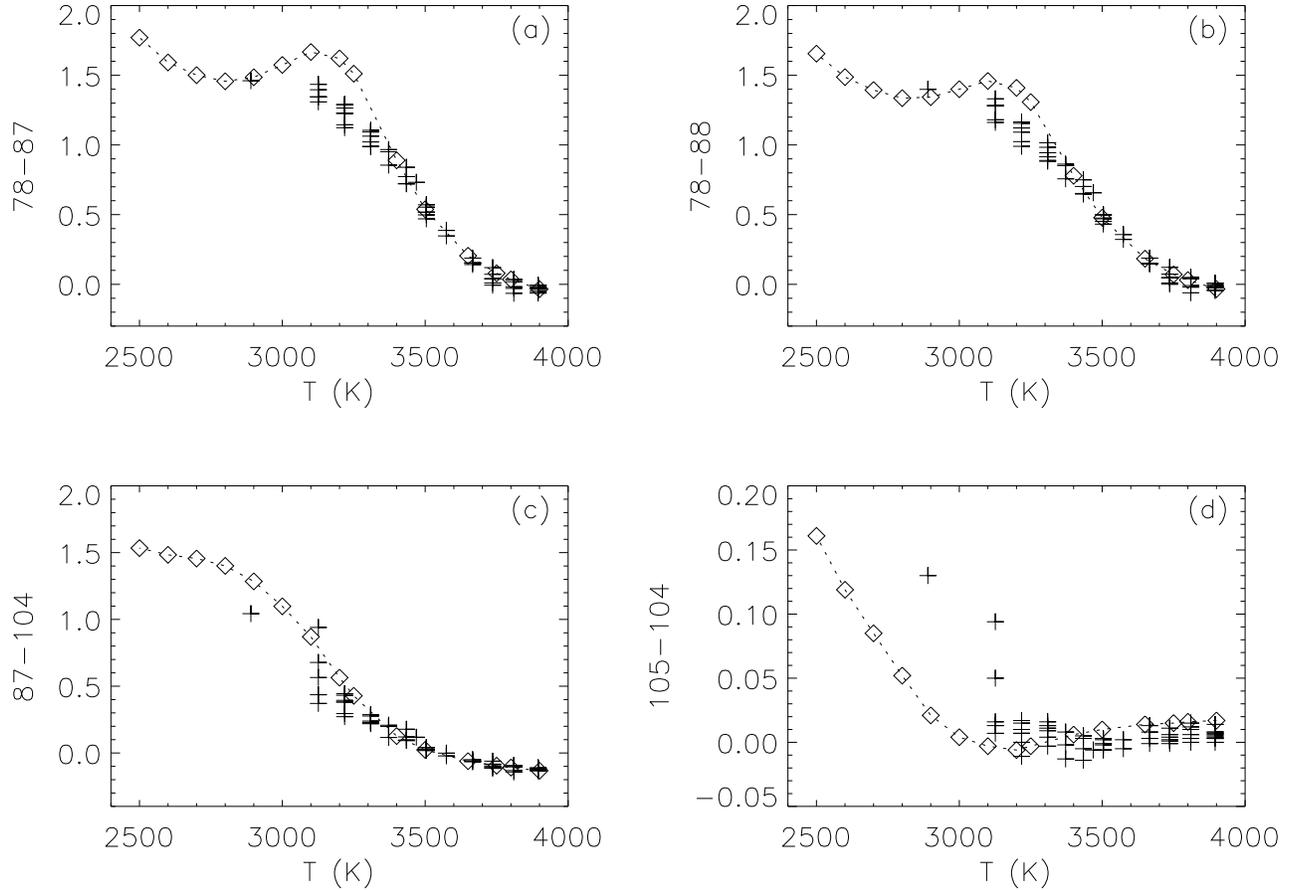,width=18cm}}
  \caption[]{Comparison of synthetic and observed indices in 
   temperature--colour diagrams. Crosses are the red giants observed 
   by Lockwood. Diamonds represent the sequence of hydrostatic models 
   with $T_{\ieff}$ ranging from 2500 to 3900~K}
\end{figure*}

A small but systematic discrepancy exists for the 78$-$88 and 78$-$87 
indices when they are greater than 0.8 (Fig~7a and~b). 
Furthermore, both colours are not ever-increasing along the sequence 
of $T_{\ieff}$.
The observed variation with phase of these indices in the Miras 
sample brings the conclusion that this behaviour is apparently not 
real.
The synthetic 87$-$104 colours are maybe also slightly above the 
observed ones for temperatures lower than 3300~K (Fig~7c). 
The 105$-$104 index appears to be the least well-modelled (Fig~7d), 
but note that its scale of variation is a factor 10 smaller than the 
other colours.\\
Discrepancies might be due to uncertainties in the:
\begin{itemize}
  \item narrow-band filter shapes (a gaussian shape was assumed) 
        and in the zero point calibration: this may imply a 
        systematic error. It seems this is the case at least for the 105 
        filter (see Fig~7d). Standard errors quoted by Lockwood for
        104 are $\pm 0.012$ mag. Errors in our zero points are 
        probably of the same order of magnitude. This is problematic 
        only for 105$-$104, which has a very small range of variation.
  \item spectral types attributed by Lockwood: some stars have the
        same spectral type whereas they exhibit rather distinct 
        colours. Furthermore, the scale of spectral subtypes (M0 to 
        M10) adopted by Fluks et al.\ for their SpT--$T_{\ieff}$ 
        relation might differ from the one used by Lockwood.
  \item laboratory measurements and calculations of lifetime as 
        quoted by the different authors: the uncertainties are still 
        important despite great progress made 
        recently in the TiO and VO bands calibration. 
\end{itemize}
Furthermore, red giants naturally exhibit a spread of effective 
temperature, but also of stellar masses, gravities and chemical
composition. 
Figure~7 shows only a 'one-dimensional' fit (in $T_{\ieff}$): $\cal M$ 
and $\log g$ have a non-negligible effect on colours and part of the
discrepancy might also be explained by this. Test calculations using
models taken from the grid of Plez (1992) show that two models with 
the same $T_{\ieff}$ and luminosity but with distinct $\log g$ and 
$\cal M$ have differences of tenths of mag for some indices (e.g.\ at 
3000~K, the $\log g$=$-0.7$ and $\cal M$=1 model differs by $-0.06$ 
in 78$-$87, $-0.007$ in 78$-$88, 0.37 in 87$-$104, and 0.042 in 
105$-$104 from the $\log g$=0.00 and $\cal M$=5 model).
We recall that the sequence of models was computed using the Ridgway 
et al.\ (1980) temperature calibration, and a theoretical red-giant 
branch from Lattanzio (1991) to relate the effective temperature and 
luminosity, assuming a mass of 1.5 $\cal M_{\odot}$.\\ 
The effect of relative abundance changes in C and N was also
checked. Some spectra were recomputed with a change of $-0.2$ dex
in the C abundance and +0.5 dex in the N abundance, which are typical
values after dredge-up (Smith 1990).  
The difference is less than a tenth of a magnitude in all colours at 
3100K, and largest for 87$-$104.\\
The fit for the 105$-$104 index is not excellent below $T_{\ieff}$
$\sim$ 3100~K. However, the variation scale is less than 0.2 mag. 
Uncertainties in the observations are of the order of 0.006 mag
for the colours, according to Lockwood. We suspect the flux level of 
our models around 1$\mu$m to be slightly too high. This appears also 
when comparing them to observed spectra of M--giants. There may be a 
missing opacity in our computations. This will be further investigated 
by Plez et al.\ (in preparation). 
Note also that the points at $T_{\ieff} \leq$ 3100~K consist of X~Her, 
EP~Aqr, BK~Vir, RT~Vir, SW~Vir and RX~Boo, which are all variables.
In conclusion, keeping in mind the above discussion, the agreement 
between the synthetic and observed colours is remarkable in Fig.~7, 
and provides a quite safe basis for the study of Miras.

\section{Miras and hydrodynamical model colours}

Now that we are confident about the line lists and our synthetic colour
calculations we may try to compute them for Mira stars.
An examination of Fig.~1 readily shows that the variation of the 
87$-$104 and 105$-$104 colours for Miras greatly exceeds the range of 
static models, whereas this is not the case for 78$-$87 and 78$-$88. 
Together with the phase lag between 105$-$104 and the other colours, 
this shows that static models will not be able to reproduce the colour 
variations along the cycle of Mira stars. 

S.\ H\"ofner (1997) kindly provided us with a sequence 
of models (50 in total distributed on 2 successive cycles) with the
parameters: $\cal M_{\star}=$1~$\cal M_{\odot}$, 
$L_{\star}=7000$~$L_{\odot}$, $T_{\star}=2880$~K, $P=390$~d, and a 
piston velocity of 4~km.s$^{-1}$. 
This model was already used by Loidl et al.\ (1997) to synthetize 
C-rich IR spectra and is similar to the models presented by 
H\"ofner \& Dorfi (1997) except for the gas opacity which was
adjusted by comparison with standard model atmospheres to give more 
realistic gas densities. 
It was computed with C/O=1.8, which is not adapted to our O-rich 
objects. However, there are no suitable O-rich hydrodynamical models 
available, and we only want to qualitatively understand the phase lag. 
A likely explanation is that this phase lag arises from the difference 
in formation depth of the various spectral regions dominated by 
continuum, TiO or VO bands, combined with the running temperature 
$T$-- and density $\rho$-- perturbations along a cycle. 
For the purpose of this work we may assume that the chemistry used in 
the computation of the hydrodynamic model is not a key issue 
(especially as it is hardly reflected in the radiative transfer, 
treated in the grey case with a constant opacity of $10^{-2}$ 
cm$^2$.g$^{-1}$; the C-rich chemistry has however a strong impact on 
the dust formation). 
We therefore only used the $T$ and $\rho$ stratification of the 
models in $R$, recomputed the chemical equilibrium with a solar 
composition, to derive the electronic pressure $P_{\ie}$ and an 
optical depth scale $\tau$. We did not use all the 500 points of the 
original models but selected about 65 depth-points with an average 
separation of 0.2 in $\Delta \log \tau$, which is sufficient for 
treating the radiative transfer.
We slightly extrapolated the models inwards to increase the maximum 
$\tau$ value in order to provide a better inner boundary condition 
(diffusion approximation), but the exact stratification deep in the 
model has no strong impact on the emergent spectrum.\\ 
As noted above, we surmise that the phase lag is due to a difference 
in the depth of formation of VO bands (dominating at 105) and TiO 
bands. A first and quick qualitative check is offered by the variation 
in column density of TiO and VO along a pulsation cycle of the model.
The column densities are obtained by calculating the chemical 
equilibrium at each depth point of the dynamical models and integrating
the number densities radially. The column densities oscillate up
and down along the cycle (Fig.~8).
The phase lag is 0.065 at minimum column density and 0.030 at maximum,
the VO variation lagging behind TiO's, providing a first qualitative
confirmation of our hypothesis. The ratio of the column 
densities is about 50, a factor 5 larger than the abundance ratio of 
Ti and V.\\

\begin{figure}
  \centerline{\psfig{figure=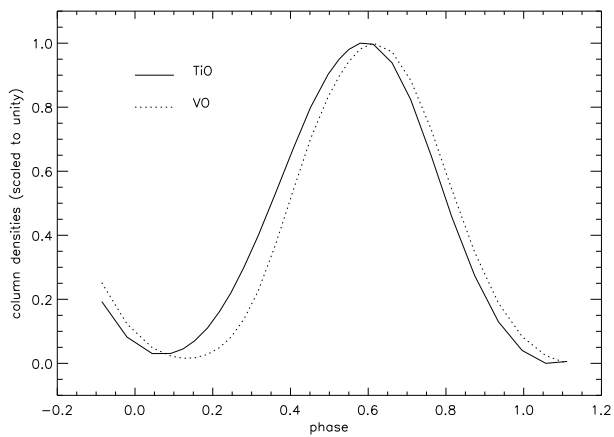,width=3.4in}}
  \caption[]{TiO and VO column densities as a function of phase,
   computed assuming chemical equilibrium in the hydrodynamical 
   model of H\"ofner. Both column density variations have been 
   scaled to the [\,0\,;\,1\,] interval to facilitate comparison. 
   The column densities vary between 3.9\,10$^{17}$ and 
   1.6\,10$^{18}$~g.cm$^{-2}$ for TiO, and 5.2\,10$^{15}$ and 
   5.0\,10$^{16}$~g.cm$^{-2}$ for VO}
\end{figure}

We also computed LTE synthetic spectra for these models in the same 
fashion as for static models. The velocity fields were neglected. 
The resulting colours are displayed in 
Fig.~9. To facilitate the comparison with observations we have also 
plotted in this figure the mean relation from Fig.~1 for each colour.
We have allowed for 2 shifts of the observed colours, relative to the
observations. The first shift, identical for all 4 colours, is in phase 
(the calculated curves are shifted by 0.08). 
There are no {\it a priori} reasons for the zero of the phase to be the 
same in the observations and in the models. For the observations, we 
recall that the zero is defined as the maximum light in V band. In the 
models, the zero is defined as the time when the piston moves outwards 
with its maximum velocity.
We also shifted the data vertically to allow an easier comparison of 
the shape and amplitude of the curves. The scale on the right-hand 
side of the diagrams is for the observed data, the one on the 
left-hand side is for the synthetic colours. The agreement is
surprisingly good. 

\begin{figure*}
  \centerline{\psfig{figure=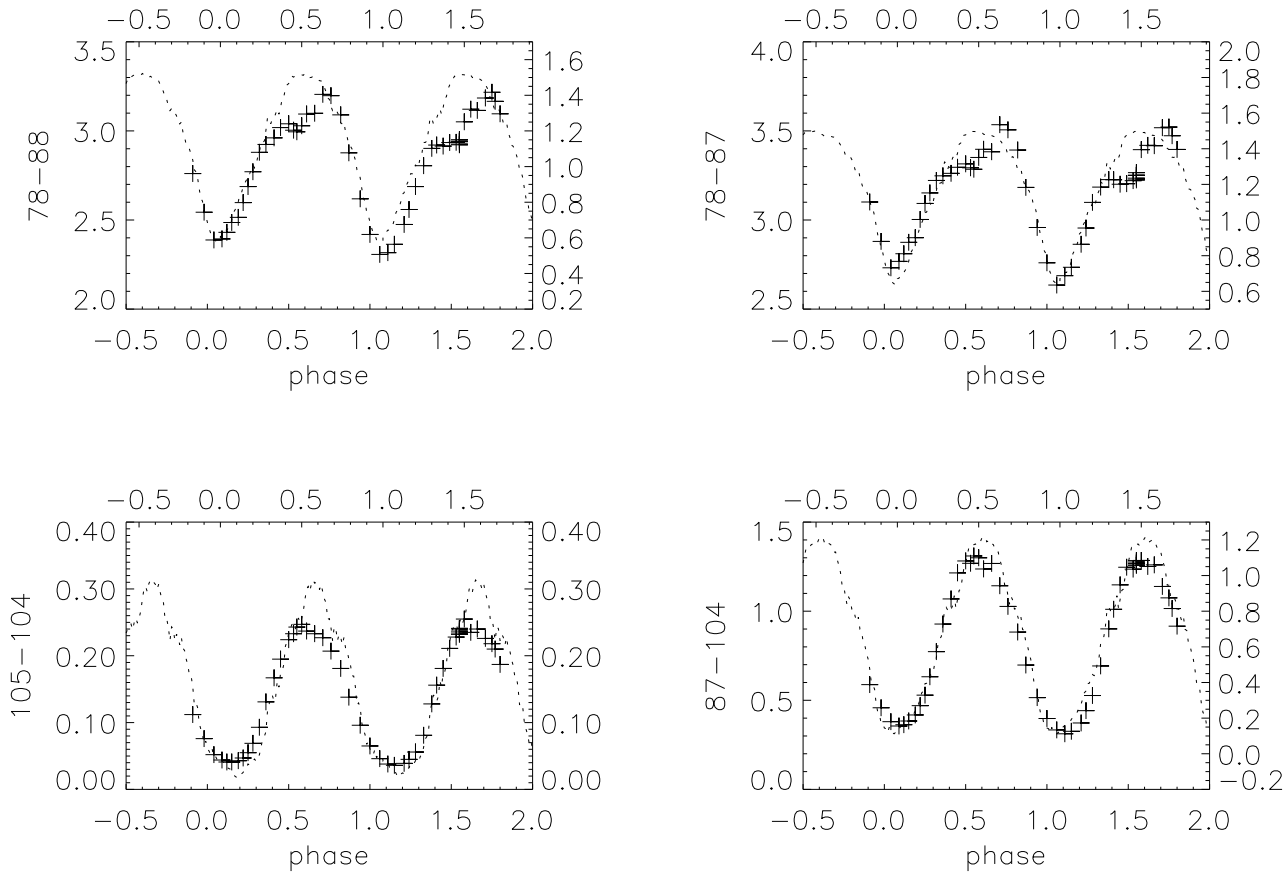,width=18cm}}
  \caption[]{Comparison of synthetic colours of dynamic models (crosses; 
   left-hand side and bottom scales) and observed mean relations from 
   Fig.~1 (dotted curves; right-hand side and top scales)}
\end{figure*}

The minima are all reasonably well matched with the possible exception
of the 87$-$104 index. From the calculated minima we find a phase lag
of 0.09 of 105$-$104 relative to 78$-$87 and 78$-$88 in very good
agreement with the value derived from observations (0.08). We also 
find a slight phase lag for 87$-$104 of 0.04, which is also detectable
in the observations at minimum light, and due to the hybrid nature of 
this index.
The maxima are more discrepant, especially for 78$-$87 and 78$-$88, 
with a tendency for the synthetic colours to show a secondary minimum. 
We suspect this to be due to the same cause as the S-shape shown in 
Fig.~7 by the synthetic colours of static models around 3000~K; we 
believe this S-shape to be in error.
It is possible that additional absorbers missing 
in our synthetic spectra would solve the discrepancy. 
A more serious problem is that the 78$-$87 and 78$-$88 synthetic 
indices are 1.5 to 2 mag too large in absolute value, although their
amplitude of variation is about right.
As 87$-$104 is of the right magnitude, it could be that this problem 
stems from an inadequate value of the 78 flux. A more detailed 
discussion is not useful given the approximations we have made. 
We have used {\it one} specific {\it C-rich} hydrodynamical model 
stratification for which we have computed an {\it LTE} radiation 
transfer with an {\it O-rich} equilibrium chemistry. We have compared 
the result of these calculations for a unique set of model parameters 
to {\it mean} colour variations resulting from the average of 256 
different Mira stars. It is actually amazing that we get such a good 
agreement. 
But one should certainly perform the same computations for a series of 
models with various $\cal M_{\star}$, $L_{\star}$, $P$, $\dot{\cal M}$, 
and check for possible variations of the phase lag, of the amplitude 
of variations of the various indices, etc. 
It would also be desirable to use O-rich models. One more limitation 
of our approach is that we have neglected the velocity field that 
shifts spectral lines differentially with depth. In this particular 
model the velocity of the gas is of the order of +15 to -15 
km.s$^{-1}$  in the line formation zone. 
This will spread out the opacity, block more flux between the lines 
and allow more photons to escape within the lines. This may solve some 
of the discrepancies we discussed above.

\section{Conclusions}

Using Lockwood's (1972) five-colours observations of 256 O-rich Miras 
we found evidence for the loops described by these stars in 
colour--colour diagrams, and a phase lag between the VO and TiO 
indices, reflecting the shocks running through the extended stellar 
atmosphere. 
We used the best current hydrostatic and hydrodynamic model 
atmospheres, and line lists to compute synthetic colours for normal 
M--giants and Miras.
We found that the normal M--giant colours are rather well reproduced, 
at least for $T_{\ieff}$ above 3100~K. Missing opacity sources, 
non-LTE, non-thermal heating, variability may all explain the worse 
match to observations at later spectral-types. This is worth further 
study, but for earlier spectral-types model atmospheres and synthetic 
spectra are reaching a high quantitative agreement with observations. 
This is very encouraging for the next (and last) generation of 
classical models in preparation in Uppsala that will include even
more up-to-date and more complete opacities (Plez et al.\ in
preparation).

Our experiments in solving LTE radiative transfer in one series
of hydrodynamical models (H\"ofner, private communication) have worked 
beyond our expectations. 
We found approximately the right variation of the colours and the 
right phase lag between them. The fit is of course not perfect, and 
should not be. We used one specific model and the observations are 
an average of 256 different stars. There are many more limitations in 
the modelling (e.g.\ LTE, no account of velocity field, ...), but the 
main point is that hydrodynamical models, equilibrium chemistry, and 
LTE radiative transfer seem to reproduce at least some of the observed 
data on Miras (cf. also the work of Loidl et al.\ 1997 on IR spectra 
of C-rich Miras). 
This may be surprising at first sight. We cannot just assume that 
equilibrium and LTE prevail all through Miras atmospheres. 
This has to be checked, and that is a heavy task. The advantage of 
the present approach is that it allows us to account for millions of 
spectral lines, carry a large number of tests (vary chemical 
composition, ...) while keeping the computation time into reasonable 
bounds. We could easily compute center-to-limb intensity profiles as 
well. We believe this work opens the prospect of calculating accurate 
colours and spectra for cool stars, not least long-period 
variables (LPV's). The out-of-phase variations of the various spectral 
features, reflecting the propagation of perturbations (shocks) in the 
atmospheres are a very strong constraint to the models.
The confrontation to spectrophotometric observations, resolving the
absorption bands, along the whole cycle, will be necessary to test the 
models and better understand LPV's atmospheres, pulsation and 
mass-loss. It will also unveil limitations/inadequacies in the 
hypotheses made in the modelling. 
This would be a very useful program for a small telescope equiped with 
an $R\approx 10\,000$ spectrograph, ideally reaching wavelengths up to 
1.1 micron.

\begin{acknowledgements}
S.\ H\"ofner and her collaborators are warmly thanked for providing us
with extensive data from one of their hydrodynamical model and for 
useful discussions and suggestions. S.\ Langhoff is acknowledged for
sending us the result of his TiO calculations prior to publication.
We thank U.G. J\o rgensen for financial support while RA was staying at the 
Niels Bohr Institute. Part of this work was carried out while BP held
a fellowship from the EU HCM program at NBI. BP thanks the GRAAL for a 
most pleasant stay during the final stage of this work.
\end{acknowledgements}


\begin{thebibliography}{}

   \bibitem{}
      Adam A.G., Barnes M., Berno B., Bower R.D., Merer A.J., 1995,
      J.\ Molec.\ Spectrosc.\ 170, 94

   \bibitem{}
      Alvarez R., Mennessier M-O., 1997,
      A\&A 317, 761

   \bibitem{}
      Bessell M.S., Brett J.M., 1988,
      PASP 100, 1143

   \bibitem{}
      Bessell M.S., Scholz M., Wood P.R., 1996,
      A\&A 307, 481

   \bibitem{}
      Brett J.M., 1990,
      A\&A 231, 440

   \bibitem{}
      Bowen G.H., 1988,
      ApJ 329, 299

   \bibitem{}
      Cheung S.C., Hansen R.C., Merer A.J., 1982a,
      J.\ Molec.\ Spectrosc.\ 91, 165

   \bibitem{}
      Cheung S.C., Taylor A.W., Merer A.J., 1982b,
      J.\ Molec.\ Spectrosc.\ 92, 391

   \bibitem{}
      Cheung S.C., Hajigeorgiou P.G., Huang G., 
      Huang S.Z., Merer A.J., 1994,
      J.\ Molec.\ Spectrosc.\ 163, 443

   \bibitem{}
      Davis S.P., Littleton J.E., Phillips J.G., 1986,
      ApJ 309, 449

   \bibitem{}
      Doverst\aa l M., Weijnitz P., 1992,
      Molec.\ Physics 75, 1375

   \bibitem{}
      Dreiling L.A., Bell R.A., 1980,
      ApJ 241, 736

   \bibitem{}
      Feinberg J., Davis S.P., 1977,
      J.\ Molec.\ Spectrosc.\ 69, 445 

   \bibitem{}
      Feuchtinger M.U., Dorfi E.A., 1994,
      A\&A 291, 209

   \bibitem{}
      Feuchtinger M.U., Dorfi E.A., 1996a,
      A\&A 306, 837

   \bibitem{}
      Feuchtinger M.U., Dorfi E.A., 1996b,
      A\&A in press

   \bibitem{}
      Fleischer A.J., Gauger A., Sedlmayr E., 1992,
      A\&A 266, 321

   \bibitem{}
      Fluks M.A., Plez B., Th\'e P.S.\ et al., 1994,
      A\&AS 105, 311

   \bibitem{}
      Gustafsson B., Bell R.A., Eriksson K., Nordlund \AA., 1975,
      A\&A 42, 407

   \bibitem{}
      Hedgecock I.M., Naulin C., Costes M., 1995,
      A\&A 304, 667

   \bibitem{}
      H\"ofner S., Dorfi E.A., 1997,
      A\&A 319, 648

   \bibitem{}
      J\o rgensen U.G., 1994,
      A\&A 284, 179

   \bibitem{}
      J\o rgensen U.G., Larsson M., 1990,
      A\&A 238, 424

   \bibitem{}
      Karlsson L., Lindgren B., Lundevall C., Sassenberg U., 1997,
      J.\ Molec.\ Spectrosc.\ 181, 274

   \bibitem{}
      Kholopov P.N.\ (ed.), 1985, 1987,
      General Catalogue of Variable Stars, Fourth Edition,
      Nauka Publ.\ House, Moscow

   \bibitem{}
      Kurucz R.L., 1992,
      Rev.\ Mex.\ Astron.\ Astrofis.\ 23, 45 

   \bibitem{}
      Langhoff S.R., 1997,
      ApJ 481, 1007

   \bibitem{}
      Larsson M., 1983,
      A\&A 128, 291

   \bibitem{}
      Lattanzio J.C., 1991,
      ApJS 76, 215

   \bibitem{}
      Lockwood G.W., 1972,
      ApJS 24, 375 (L72)

   \bibitem{}
      Loidl R., Hron J., H\"ofner S.\ et al., 1997,
      ApSS in press 

   \bibitem{}
      Plez B., 1992,
      A\&AS 94, 527

   \bibitem{}
      Plez B., 1997, 
      In: R.F.\ Wing (ed.) The Carbon Star Phenomenon, in press

   \bibitem{}
      Plez B., Brett J.M., Nordlund \AA., 1992,
      A\&A 256, 551 (PBN92)

   \bibitem{}
      Plez B., Smith V.V., Lambert D.L., 1993,
      ApJ 418, 812

   \bibitem{}
      Ramsey L.W., 1981
      AJ 86, 557

   \bibitem{}
      Ridgway S.T., Joyce R.R., White N.M., Wing R.F., 1980,
      ApJ 235, 126   

   \bibitem{}
      Schamps J., Sennesal J.M., Carette P., 1992,
      J.\ Quant.\ Spectrosc.\ Rad.\ Transfer 48, 147

   \bibitem{}
      Serote Roos M., Boisson C., Joly M., 1996
      A\&AS 117, 93

   \bibitem{}
      Smith V.V., 1990, 
      Mem.\ Soc.\ Astron.\ Ital.\ 61, 787

   \bibitem{}
      Wing R.F., 1967,
      In: M.\ Hack (ed.) Colloquium on late-type stars,
      Osservatorio Astronomico di Trieste, p.231

   \bibitem{}
      Ya'ari A., Tuchman Y., 1996,
      ApJ 456, 350

\end{thebibliography}
\end{document}